\documentclass[prd,twocolumn,showpacs,floatfix,amsmath,amssymb,floatfix]{revtex4}
\usepackage{graphicx,dcolumn,booktabs,bm}
\usepackage{longtable,lscape}
\usepackage{txfonts}
\usepackage{overpic}
\usepackage{multirow}
\usepackage{amssymb}
\usepackage{indentfirst}
\usepackage{feynmf}   
\usepackage{slashed}  
\usepackage{cases}
\usepackage{color,ulem}
\usepackage{graphicx}

\graphicspath{{Figures/}} 

\begin{document}


\title{Exotic Four Quark Matter: $Z_1(4475)$}
\author{Li Ma$^{1}$}\email{lima@pku.edu.cn}
\author{Xiao-Hai Liu$^{1}$}\email{liuxiaohai@pku.edu.cn}
\author{Xiang Liu$^{2,3}$}\email{xiangliu@lzu.edu.cn}
\author{Shi-Lin Zhu$^{1,4}$}\email{zhusl@pku.edu.cn}

\affiliation{
$^1$Department of Physics and State Key Laboratory of Nuclear Physics and Technology and Center of High Energy Physics, Peking University, Beijing 100871, China\\
$^2$Research Center for Hadron and CSR Physics, Lanzhou University and Institute of Modern Physics of CAS, Lanzhou 730000, China\\
$^3$School of Physical Science and Technology, Lanzhou University, Lanzhou 730000, China\\
$^4$Collaborative Innovation Center of Quantum Matter, Beijing
100871, China}

\date{\today}

\begin{abstract}

Motivated by the LHCb's recent confirmation of $Z_1(4475)$ as the
$J^P=1^+$ resonance, we investigate various exotic interpretations
of $Z_1(4475)$, which may be an axial vector tetraquark state, the
P-wave excitation of the S-wave $D_1 {\bar D}^\ast$ or $D_2 {\bar
D}^\ast$ molecule, the S-wave molecule composed of a $D$ or $D^*$
meson and a D-wave vector $D$ meson, or the cousin molecular state
of $Z_c(3900)$ and $Z_c(4020)$ composed of a $D$ or $D^*$ meson and
their radial excitations. With the help of the heavy quark symmetry,
we predict the typical radiative and hidden-charm and open-charm
strong decay patterns of $Z_1(4475)$, which are crucial to further
identify the molecular state assignment of $Z_1(4475)$.

\end{abstract}

\pacs{14.40.Rt, 13.20.Jf, 13.25.Jx} \maketitle

\section{Introduction}\label{sec1}

As a great achievement in the understanding of matter, Quantum
Chromodynamics (QCD) is established as the theory of the strong
interaction. However, QCD is highly non-perturbative in the infrared
region and the color confinement issue remains one of the most
challenging problems in 21st-century science. It is well known
that a nucleon is composed of three quarks and a meson is composed
of a pair of quark and anti-quark. Besides the above conventional
hadrons, QCD may allow the existence of the subatomic particles with
the other quark/gluon configurations such as hybrid hadrons,
hadronic molecules or tetraquarks etc. The experimental and
theoretical investigations of these exotic states will shed light on
the non-perturbative dynamics of QCD.

Very recently, the LHCb experiment has confirmed the existence of
the exotic hadrons \cite{Aaij:2014jqa}. $Z_1(4475)$ is the first
charged charmonium-like state discovered by the Belle Collaboration
in the $\psi' \pi$ mode in the $B$ decays \cite{Choi:2007wga}.
Later, the measurement indicates that the spin parity $1^+$ was
favored over other assignments $0^-$, $1^-$, $2^-$ and $2^+$ by
$3.4\sigma$, $3.7\sigma$, $4.7\sigma$ and $5.1\sigma$, respectively
\cite{Chilikin:2013tch}.
Recently, the LHCb Collaborations confirmed $Z_1(4475)$ as the
$J^P=1^+$ state with a significance of $13.9\sigma$ and ruled out
the other quantum numbers with less significance
\cite{Aaij:2014jqa}. Its mass and width are measured to be $4475$
MeV and $172$ MeV, respectively. From now on, we will denote this
charged state as $Z_1(4475)$. The obtained Argand diagram for the
$Z_1(4475)$ amplitude is consistent with the resonance behavior.
Moreover, LHCb observed a second charged state $Z_0(4239)$ with
$J^P=0^-$ with a significance of $6\sigma$ and a width around $220$
MeV.

Since its observation, $Z_1(4475)$ was first considered to be a good
candidate of the loosely bound S-wave molecular state composed of
the $D_1^{(\prime)} {\bar D}^\ast$ meson pair
\cite{Meng:2007fu,Liu:2007bf,Ding:2007ar,Liu:2008xz,Liu:2008yy,Lee:2007gs}. In particular, it was pointed out in
Ref. \cite{Liu:2008xz} that there exists only the $J^P=0^-$ molecule for
the S-wave $D_1 {\bar D}^\ast$ system. However, there exist the
S-wave $D_1^\prime {\bar D}^\ast$ molecule with $J^P=0^-, 1^-, 2^-$
\cite{Liu:2008xz}. The authors pointed out that the broad width of
$D_1^\prime$ seems not very compatible with the narrow width of
$Z_1(4475)$ around $45$ MeV measured in Belle's discovery paper
\cite{Choi:2007wga}.
LHCb's precise measurement shows that $Z_1(4475)$ sits exactly on
the threshold of $D_2^*(2460)$ and ${\bar D}^*(2010)$ and is
slightly above the $D_1(2420) {\bar D}^\ast$ threshold. The
confirmation of its spin parity as $J^P=1^+$ leads to very puzzling
new challenges, which are also a good opportunity to demystify $Z_1(4475)$ as the
four quark matter.

There are several interesting schemes of the underlying structure of $Z_1(4475)$.
The first possibility is that $Z_1(4475)$ may be the $J^P=1^+$
hidden-charm tetraquark candidate. The axial-vector charmonium-like
tetraquark mass spectrum are discussed extensively in Ref.
\cite{Chen:2010ze}. As a tetraquark candidate, $Z_1(4475)$ should decay
into the $J/\psi \pi$ mode more easily with a larger partial width
than that of the discovery mode $\psi'\pi$. Moreover, $Z_1(4475)$
should also decay into the open-charm modes ${\bar D}D^*$ and ${\bar
D}^* D^*$ very easily via S-wave. Then, $Z_1(4475)$ should be a very
broad state with the total width much larger than $172$ MeV, which
seems not compatible with the Belle and LHCb's precise measurement
of its width.

The second possibility is that $Z_1(4475)$ is the P-wave excitation
of the S-wave $D_1 {\bar D}^\ast$ or $D_2 {\bar D}^\ast$ molecule.
If so, is $Z_0(4239)$ the ground state of the $D_1 {\bar D}^\ast$
molecule? Then, the binding energy of $Z_0(4239)$ is as large as
$190$ MeV, which is in strong contrast with the tiny binding energy
(around 2 MeV) of the deuteron. Generally one would expect that the
strong interaction between two color-singlet hadrons leads to a
binding energy around a few to several tens MeV instead of hundreds
of MeV, although the very deeply bound hadronic molecules may also
exist.

The third possibility is that $Z_1(4475)$ is the S-wave molecule
composed of a $D$ or $D^*$ meson and a D-wave vector $D$ meson in
the $(1^-, 2^-)$ multiplet whose mass is around 2700 MeV.

The fourth possibility is that $Z_1(4475)$ is the S-wave molecule
composed of a $D$ or $D^*$ meson and their radial excitations whose
masses are around 2.6 GeV. In other words, $Z_1(4475)$ may be the
cousin of the charged states $Z_c(3900)$ and $Z_c(4025)$ observed by the
BESIII Collaborations \cite{bes,Liu:2013dau}, which are speculated to be
molecular candidates composed of the $D$ and $D^*$ mesons.

The discovery mode $\psi'\pi$ and non-observation of $Z_1(4475)$ in
the $J/\psi \pi$ channel are also serious challenges to the second
and third possibilities listed above. However, the situation is
quite different for the last scheme. If $Z_1(4475)$ contains the
radial excitation of the $D$ or $D^*$ meson as its component, it may
decay into the final state containing a radial excitation more
easily, i.e., the $\psi'\pi$ final state may be a more favorable
decay mode of $Z_1(4475)$ than $J/\psi \pi$ within the fourth
scheme.

In this work we will investigate the implications of the Belle and
LHCb's determination of the spin parity of $Z_1(4475)$. With the
help of heavy quark symmetry, we will study the electromagnetic,
both hidden-charm and open-charm strong decay patterns of
$Z_1(4475)$ under various molecular assumptions. Some interesting
features can be tested experimentally.

This paper is organized as follows. After the Introduction, we present the formalism and results in Sec.
\ref{sec2}. The last section is a short summary.

\section{The decay behavior of $Z_1(4475)$}\label{sec2}

The quantum number of the neutral component of $Z_1(4475)$ is
$J^{PC}=1^{+-}$. As a molecular state candidate, the possible flavor
wave functions of $Z_1(4475)$ are listed in Table \ref{state}. Here, $D(\frac{3}{2},1)$  is a D-wave charmed meson with the light spin $S_l=\frac{3}{2}$ and $J^P=1^-$.
$D(2550)$ and $D^\ast(2600)$ were reported by BaBar \cite{delAmoSanchez:2010vq}, which can be good candidates of $2^1S_0$ and $2^3S_1$ charmed  mesons, respectively \cite{Sun:2010pg}. $Z_1(4475)$ can
be a P-wave $D_1\bar{D}^\ast$ or $D_1'\bar{D}^\ast$ molecular state.
$Z_1(4475)$ may also be an S-wave molecular state containing a D-wave
meson or a radially excited state.
\renewcommand{\arraystretch}{1.4}
\begin{table}[htbp]
\caption{The flavor wave functions of $Z_1(4475)$ as a molecular
candidate.\label{state}}
\begin{center}
   \begin{tabular}{c|ccc } \toprule[1pt]
    &\multicolumn{3}{c}{States}\\
   \midrule[1pt]

     \multirow{8}{*}{P-wave molecules} & \multirow{2}{*}{$^1P_1$} & $\frac{1}{\sqrt{2}}(D_1'\bar{D}^\ast-D^\ast\bar{D}_1')$ \\
     &&$\frac{1}{\sqrt{2}}(D_1\bar{D}^\ast-D^\ast\bar{D}_1)$  \\\cline{2-3}

        & \multirow{3}{*}{$^3P_1$} & $\frac{1}{\sqrt{2}}(D_1'\bar{D}^\ast+D^\ast\bar{D}_1')$
        \\&&$\frac{1}{\sqrt{2}}(D_1\bar{D}^\ast+D^\ast\bar{D}_1)$  \\

        & & $\frac{1}{\sqrt{2}}(D_2\bar{D}^\ast-D^\ast\bar{D}_2)$ &  \\\cline{2-3}

        & \multirow{3}{*}{$^5P_1$} & $\frac{1}{\sqrt{2}}(D_1'\bar{D}^\ast-D^\ast\bar{D}_1')$ \\&&$\frac{1}{\sqrt{2}}(D_1\bar{D}^\ast-D^\ast\bar{D}_1)$  \\

        & & $\frac{1}{\sqrt{2}}(D_2\bar{D}^\ast+D^\ast\bar{D}_2)$ & \\\midrule[1pt]

      \multirow{4}{*}{S-wave molecules} & &  $\frac{1}{\sqrt{2}}(D(\frac{3}{2},1)\bar{D}-D\bar{D}(\frac{3}{2},1))$ \\&& $\frac{1}{\sqrt{2}}(D(\frac{3}{2},1)\bar{D}^\ast+D^\ast\bar{D}(\frac{3}{2},1))$ \\

      & & $\frac{1}{\sqrt{2}}(D(2550)\bar{D}^\ast-D^\ast\bar{D}(2550))$\\&
      & $\frac{1}{\sqrt{2}}(D\bar{D}^\ast(2600)-D^\ast(2600)\bar{D})$ \\
      \bottomrule[1pt]
       \end{tabular}
 \end{center}
\end{table}

The heavy quark spin symmetry \cite{Manohar:2000dt} is very useful in the study of the heavy
hadron properties. In the heavy quark limit, the heavy quark mass
$m_Q\rightarrow\infty$, the spin-flipping interaction is suppressed.
Throughout the decay process, the heavy quark spin $S_H$ is a
conserved quantity, which is named as "heavy spin" for simplicity.
The total angular momentum $J$ of a hadron is also a conserved
quantity. We can also define the conserved "light spin"
$\vec{S}_l\equiv \vec{J}-\vec{S}_H$, which includes both the light
quark spin and the orbital angular momenta within a hadron.
In the decays of a hadron, the heavy spin, light spin and total
angular momentum are all good quantum numbers in the heavy quark
limit, which are separately conserved. Therefore we can decompose
the total angular momentum of the initial and final states according
to their heavy spin and light spin, which was employed to study the
radiative decays of $XYZ$ states extensively in Ref. \cite{Ma:2014ofa}. We
borrow the same notations and formalism from Ref. \cite{Ma:2014ofa} to study
the radiative and strong decay patterns of $Z_1(4475)$.
Since the charm quark is not very large, the heavy quark symmetry is
not exact. In the following, we analyze the hidden-charm molecular
systems in the heavy quark symmetry limit. One may naively expect
the recoil corrections from the finite charm quark mass will not
spoil the qualitative features outlined in this work. We focus on
the P-wave $D_1\bar{D}^\ast$ or $D_1'\bar{D}^\ast$, S-wave
$D(\frac{3}{2},1)\bar{D}$ or $D(\frac{3}{2},1)\bar{D}^\ast$, as well
as S-wave $D(2550)\bar{D}^\ast$ or $D\bar{D}^\ast(2600)$ molecular
systems in this work.

With the heavy quark spin symmetry \cite{Manohar:2000dt}, the heavy and light spins of the
molecular state can be re-coupled separately. We adopt the spin
re-coupling formula with 6-$j$ or 9-$j$ symbols in analyzing the
general spin structure. In the heavy quark limit, the final state charmonia can also be decomposed into the heavy spin and light spin. For the neutral component of $Z_1(4475)$, there exists the radiative
decay $Z_1(4475)\to (c\bar{c})+\gamma$, where the initial state is a
hadronic molecule and final state is a charmonium. The photon is
from the $q\bar{q}$ annihilation. We can calculate the rearranged spin
structures of the final states in the $Z_1(4475)\to
(c\bar{c})+\gamma$ decays. The general expression is \cite{Ma:2014ofa}
\begin{eqnarray}
 && |\textsf{Charmionia}\rangle\otimes|\gamma\rangle\nonumber\\
 & &= \left[[(c\bar{c})_g\otimes L ]_K\otimes Q\right]_J |(c\bar{c})\rangle|\gamma\rangle  \nonumber\nonumber\\
  &&= \sum_{h=|L-Q|}^{L+Q} (-1)^{g+L+Q+J}\Big[(2K+1)(2h+1)\Big]^{1/2}\nonumber\\
  &&\quad\times
  \left\{
    \begin{array}{ccc}
      L & g & K \nonumber\\
      J & Q & h \nonumber\\
    \end{array}
  \right\} \Bigg|\left[\left(c\bar{c}\right)_g\otimes\left[L\otimes Q\right]_h\right]_J \Bigg\rangle |(c\bar{c})\rangle|\gamma\rangle,
   \end{eqnarray}
where the $g$ and $L$ denote the heavy and light spins of the
charmonium, respectively. $Q$ stands for the light spin of the
photon. The indices $c$, $\bar{c}$ and $\gamma$ in the square
brackets represent the corresponding spin wave functions. Similarly, we present the rearranged spin
structures of the final states in the $Z_1(4475)\to
(c\bar{c})+\textsf{light}\,\, \textsf{meson}$ decays \cite{Ma:2014ofa}, i.e.,
\begin{eqnarray}
 && |\textsf{Charmionia}\rangle\otimes|\textsf{light}\,\, \textsf{meson}\rangle\nonumber\\
 & &= \left[[(c\bar{c})_g\otimes L ]_K\otimes Q\right]_J |(c\bar{c})\rangle|(q\bar{q})\rangle  \nonumber\nonumber\\
  &&= \sum_{h=|L-Q|}^{L+Q} (-1)^{g+L+Q+J}\Big[(2K+1)(2h+1)\Big]^{1/2}\nonumber\\
  &&\quad\times
  \left\{
    \begin{array}{ccc}
      L & g & K \nonumber\\
      J & Q & h \nonumber\\
    \end{array}
  \right\} \Bigg|\left[\left(c\bar{c}\right)_g\otimes\left[L\otimes Q\right]_h\right]_J \Bigg\rangle |(c\bar{c})\rangle|(q\bar{q})\rangle,
   \end{eqnarray}
With the above preparation, we obtain the typical decay behavior of
of $Z_1(4475)$ under different molecular state assignments, where
these properties follow from the heavy quark symmetry and the
presumed nature without employing a particular model, which is
crucial to test and establish the exotic four quark matter.

\subsection{$Z_1(4475)$ as a P-wave molecular candidate}
The decay pattern of $Z_1(4475)$ as a candidate of the $^3P_1$ or
$^5P_1$ $D_2\bar{D}^\ast$ molecular state is very similar to that of
the $^3P_1$ or $^5P_1$ $D_1\bar{D}^\ast$ state. We focus on the
latter. As a P-wave $D_1\bar{D}^\ast$ or $D_1'\bar{D}^\ast$
molecular state, $Z_1(4475)$ has three possible spin structures,
corresponding to $^1P_1$, $^3P_1$ and $^5P_1$, respectively. Its
neutral component can decay into $\chi_{cJ}$ $(J=0,1,2)$ via the $M1$
transition. Except the $^3P_1$ state
 $\frac{1}{\sqrt{2}}(D_1'\bar{D}^\ast+D^\ast\bar{D}_1')$, the other
five $D_1\bar{D}^\ast$ or $D_1'\bar{D}^\ast$ molecular states share
the same reduced matrix elements for the $M1$ transition. These reduced matrix elements depend on the spin
configuration $(1_H^-\otimes 0_l^-)_{J=1}^{+-}$, $(1_H^-\otimes
1_l^-)_{J=1}^{+-}$ and $(1_H^-\otimes 2_l^-)_{J=1}^{+-}$. 

We notice that both the $^3P_1$ states
$D_1'\bar{D}^\ast$ and $D^\ast\bar{D}_1'$ molecular states with
$R=1$ contain the spin configurations $(0_H^-\otimes
1_l^-)_{J=1}^{++}$, $(0_H^-\otimes 1_l^-)_{J=1}^{+-}$ and
$(1_H^-\otimes 1_l^-)_{J=1}^{++}$. The
recoupled final state $\chi_{cJ}(1^3P_J)\gamma(M1)$ with the total
angular momentum equal to $1$ contains the component of
$(1_H^-\otimes 0_l^-)_{J=1}^{+-}$, $(1_H^-\otimes 1_l^-)_{J=1}^{+-}$
and $(1_H^-\otimes 2_l^-)_{J=1}^{+-}$. So both $D_1'\bar{D}^\ast$
and $D^\ast\bar{D}_1'$ components with $R=1$ can independently decay
into $\chi_{cJ}(1^3P_J)\gamma(M1)$. Unfortunately, these two
components have the opposite relative phase for the component of
$(1_H^-\otimes 1_l^-)_{J=1}^{+-}$. When they constitute the $C$-parity
eigenstate $\frac{1}{\sqrt{2}}(D_1'\bar{D}^\ast+D^\ast\bar{D}_1')$,
the radiative decay into $\chi_{cJ}(1^3P_J)\gamma(M1)$ is suppressed
in the heavy quark limit.

The discussed $E1$ transitions include $\eta_c\gamma(E1)$ and
$\eta_{c2}(1^1D_2)\gamma(E1)$. The states
$\frac{1}{\sqrt{2}}(D_1\bar{D}^\ast-D^\ast\bar{D}_1)(^1P_1)$ and
$\frac{1}{\sqrt{2}}(D_1'\bar{D}^\ast-D^\ast\bar{D}_1')(^5P_1)$ don't
contain the spin configuration $(0_H^-\otimes 1_l^-)_{J=1}^{+-}$.
These $\eta_c\gamma(E1)$ and $\eta_{c2}(1^1D_2)\gamma(E1)$ modes are
suppressed due to heavy quark symmetry. All the other configurations
of $Z_1(4475)$ can decay into $\eta_c\gamma(E1)$ and
$\eta_{c2}(1^1D_2)\gamma(E1)$ through the spin configuration
$(0_H^-\otimes 1_l^-)_{J=1}^{+-}$.

The $E2$ decay of
$\frac{1}{\sqrt{2}}(D_1'\bar{D}^\ast+D^\ast\bar{D}_1')(^3P_1)$ is
suppressed in the heavy quark symmetry limit. Both the
$D_1'\bar{D}^\ast$ and $D^\ast\bar{D}_1'$ molecular states with
$J=1$ contain three spin configurations $(0_H^-\otimes
1_l^-)_{J=1}^{++}$, $(0_H^-\otimes 1_l^-)_{J=1}^{+-}$ and
$(1_H^-\otimes 1_l^-)_{J=1}^{+-}$. The
spin-rearranged final states $\chi_{c1}(1^3P_1)\gamma(E2)$ and
$\chi_{c2}(1^3P_2)\gamma(E2)$ contain the components $(1_H^-\otimes
1_l^-)_{J=1}^{+-}$ and $(1_H^-\otimes 2_l^-)_{J=1}^{+-}$. Both
$D_1'\bar{D}^\ast$ and $D^\ast\bar{D}_1'$ with $R=1$ can
independently decay into $\chi_{c1}(1^3P_1)\gamma(E2)$ or
$\chi_{c2}(1^3P_2)\gamma(E2)$. But these two contributions have the
opposite relative phase for the component $(1_H^-\otimes
1_l^-)_{J=1}^{+-}$. When they constitute the $C$-parity eigenstate
$\frac{1}{\sqrt{2}}(D_1'\bar{D}^\ast+D^\ast\bar{D}_1')(^3P_1)$, the
radiative decay into $\chi_{c1}(1^3P_1)\gamma(E2)$ or
$\chi_{c2}(1^3P_2)\gamma(E2)$ is suppressed in the heavy quark
limit. All the other states can
decay into $\chi_{c1}(1^3P_1)$ or $\chi_{c2}(1^3P_2)$ via the $E2$
transition.

\subsection{$Z_1(4475)$ as a S-wave molecular candidate}

The $M1$ decays of both
$\frac{1}{\sqrt{2}}(D(\frac{3}{2},1)\bar{D}-D\bar{D}(\frac{3}{2},1))$
and
$\frac{1}{\sqrt{2}}(D(\frac{3}{2},1)\bar{D}^\ast+D^\ast\bar{D}(\frac{3}{2},1))$
state occur through the spin configuration $(1_H^-\otimes
0_l^-)_{J=1}^{+-}$.
 If $Z_1(4475)$ is the
$\frac{1}{\sqrt{2}}(D(2550)\bar{D}^\ast-D^\ast\bar{D}(2550))$ or
$\frac{1}{\sqrt{2}}(D\bar{D}^\ast(2600)-D^\ast(2600)\bar{D})$
molecule, their $M1$ transitions depend on the spin configuration
$(1_H^-\otimes 0_l^-)_{J=1}^{+-}$. 
The S-wave states
$\frac{1}{\sqrt{2}}(D(2550)\bar{D}^\ast-D^\ast\bar{D}(2550))$ and
$\frac{1}{\sqrt{2}}(D\bar{D}^\ast(2600)-D^\ast(2600)\bar{D})$ do not
contain the spin configuration $(1_H^-\otimes 1_l^-)_{J=1}^{+-}$.
Thus, their $E2$ decays into $\chi_{c1}(1^3P_1)\gamma(E2)$ and
$\chi_{c2}(1^3P_2)\gamma(E2)$ are suppressed in the heavy quark
symmetry limit.


We obtain some typical $M1$ transition ratios
$\Gamma(\chi_{c0}\gamma(M1)):\Gamma(\chi_{c1}\gamma(M1)):\Gamma(\chi_{c2}\gamma(M1))$
and $\Gamma(\chi_{c1}\gamma(E2)):\Gamma(\chi_{c2}\gamma(E2))$. If $Z_1(4475)$ is an S-wave molecular state
$\frac{1}{\sqrt{2}}(D(\frac{3}{2},1)\bar{D}-D\bar{D}(\frac{3}{2},1))$
or
$\frac{1}{\sqrt{2}}(D(\frac{3}{2},1)\bar{D}^\ast+D^\ast\bar{D}(\frac{3}{2},1))$,
the above $M1$ decay width ratio is $20:15:1$ ignoring the phase space
difference. These decays are only governed by the spin configuration
$(1_H^-\otimes 2_l^-)|_{J=1}^{+-}$.

If $Z_1(4475)$ is the S-wave molecular candidate with the
configurations
$\frac{1}{\sqrt{2}}(D(2550)\bar{D}^\ast-D^\ast\bar{D}(2550))$ and
$\frac{1}{\sqrt{2}}(D\bar{D}^\ast(2600)-D^\ast(2600)\bar{D})$, the
above $M1$ decay width ratio is $1:3:5$.

\section{Strong decays of $Z_1(4475)$}\label{sec4}

$G$-parity conservation constrains the hidden-charm decay modes of
$Z_1(4475)$: $J/\psi \pi$ ($\psi'\pi$), $\psi(1^3D_1)\pi$,
$\eta_c\rho$ and $\eta_{c2}(1^1D_2)\rho$. Since the radial
excitation has the same spin decomposition as the ground state,
their decay patterns are the same. For example, it's understood
throughout our discussions that the $J/\psi\pi$ and $\psi'\pi$ modes
are the same in our symmetry analysis.

For the $^1P_1$ case, all the four hidden-charm decay modes are
allowed if $Z_1(4475)$ is a $D_1'\bar{D}^\ast$ molecule. For example, the $\eta_c\rho$ and
$\eta_{c2}(1^1D_2)\rho$ modes arise from the spin configuration
$(0_H^-\otimes 1_l^-)_{J=1}^{+-}$. In contrast, the
$\frac{1}{\sqrt{2}}(D_1\bar{D}^\ast-D^\ast\bar{D}_1)(^1P_1)$
molecular state doesn't contain the spin configuration
$(0_H^-\otimes 1_l^-)_{J=1}^{+-}$.
Hence, its decays into $\eta_c\rho$ and $\eta_{c2}(1^1D_2)\rho$ are
suppressed. Both the $D_1'\bar{D}^\ast (^1P_1)$ and
$D_1'\bar{D}^\ast (^1P_1)$ states allow the decay modes $J/\psi\pi$
and $\psi(1^3D_1)\pi$, which depend on the spin configurations
$(1_H^-\otimes 0_l^-)_{J=1}^{+-}$ and $(1_H^-\otimes
2_l^-)_{J=1}^{+-}$ respectively.

Since neither $D_1'\bar{D}^\ast$ nor $D^\ast\bar{D}_1'$ in the
$^3P_1$ state contains the component $(1_H^-\otimes
0_l^-)_{J=1}^{+-}$, the decays of
$\frac{1}{\sqrt{2}}(D_1'\bar{D}^\ast-D^\ast\bar{D}_1')(^3P_1)$ into
$J/\psi\pi$ and $\psi(1^3D_1)\pi$ are suppressed. For the
$\frac{1}{\sqrt{2}}(D_1\bar{D}^\ast-D^\ast\bar{D}_1)(^3P_1)$ state,
both $J/\psi\pi$ and $\psi(1^3D_1)\pi$ are allowed while they depend
on the $(1_H^-\otimes 0_l^-)_{J=1}^{+-}$ and $(1_H^-\otimes
2_l^-)_{J=1}^{+-}$ components, respectively. Similar conclusions hold
for the $^3P_1$ $D_2\bar{D}^\ast$ molecular state. Both
$\frac{1}{\sqrt{2}}(D_1'\bar{D}^\ast-D^\ast\bar{D}_1')(^3P_1)$ and
$\frac{1}{\sqrt{2}}(D_1\bar{D}^\ast-D^\ast\bar{D}_1)(^3P_1)$ can
decay into $\eta_c\rho$ and $\eta_{c2}(1^1D_2)\rho$ through the spin
configuration $(0_H^-\otimes 1_l^-)_{J=1}^{+-}$.

For $\frac{1}{\sqrt{2}}(D_1'\bar{D}^\ast-D^\ast\bar{D}_1')(^5P_1)$,
$\frac{1}{\sqrt{2}}(D_1\bar{D}^\ast-D^\ast\bar{D}_1)(^5P_1)$ and
$\frac{1}{\sqrt{2}}(D_2\bar{D}^\ast+D^\ast\bar{D}_2)(^5P_1)$, their
$J/\psi\pi$ and $\psi(1^3D_1)\pi$ modes are similar to the $^1P_1$
case. However, these $^5P_1$ states can also decay into $\eta_c\rho$
and $\eta_{c2}(1^1D_2)\rho$. Since neither $D_1'\bar{D}^\ast$ nor
$D^\ast\bar{D}_1'$ contains the component $(0_H^-\otimes
1_l^-)_{J=1}^{+-}$, the $\eta_c\rho$ and $\eta_{c2}(1^1D_2)\rho$
modes of
$\frac{1}{\sqrt{2}}(D_1'\bar{D}^\ast-D^\ast\bar{D}_1')(^5P_1)$ are
suppressed in the heavy quark symmetry limit.

The S-wave states
$\frac{1}{\sqrt{2}}(D(\frac{3}{2},1)\bar{D}-D\bar{D}(\frac{3}{2},1))$
and
$\frac{1}{\sqrt{2}}(D(\frac{3}{2},1)\bar{D}^\ast+D^\ast\bar{D}(\frac{3}{2},1))$,
can decay into $\psi(1^3D_1)\pi$, $\eta_c\rho$ or
$\eta_{c2}(1^1D_2)\rho$ through the components $(1_H^-\otimes
2_l^-)_{J=1}^{+-}$, $(0_H^-\otimes 1_l^-)_{J=1}^{+-}$ and
$(0_H^-\otimes 1_l^-)_{J=1}^{+-}$ respectively. However, neither
$\frac{1}{\sqrt{2}}(D(\frac{3}{2},1)\bar{D}-D\bar{D}(\frac{3}{2},1))$
nor
$\frac{1}{\sqrt{2}}(D(\frac{3}{2},1)\bar{D}^\ast+D^\ast\bar{D}(\frac{3}{2},1))$
contains the spin configuration $(1_H^-\otimes 0_l^-)_{J=1}^{+-}$.
Therefore, the $J/\psi\pi$ or $\psi'\pi$ modes are suppressed in the
heavy quark symmetry limit.

The S-wave states
$\frac{1}{\sqrt{2}}(D(2550)\bar{D}^\ast-D^\ast\bar{D}(2550))$ and
$\frac{1}{\sqrt{2}}(D\bar{D}^\ast(2600)-D^\ast(2600)\bar{D})$ can
decay into $J/\psi\pi$, $\eta_c\rho$ or $\eta_{c2}(1^1D_2)\rho$.
However, neither
$\frac{1}{\sqrt{2}}(D(2550)\bar{D}^\ast-D^\ast\bar{D}(2550))$ nor
$\frac{1}{\sqrt{2}}(D\bar{D}^\ast(2600)-D^\ast(2600)\bar{D})$
contains the spin configuration $(1_H^-\otimes 2_l^-)_{J=1}^{+-}$.
The $\psi(1^3D_1)\pi$ mode is suppressed.

In addition, we notice that all the $D_1'\bar{D}^\ast(^1P_1, ^3P_1,
^5P_1)$ and $D_1\bar{D}^\ast(^1P_1, ^3P_1, ^5P_1)$ states can decay
into ${\bar D} D^*$ and ${\bar D}^* D^*$ through either S-wave or
D-wave. The S-wave molecule containing the radiation excitation can
decay into the ${\bar D} D^*$ and ${\bar D}^* D^*$ modes. Thus,
$e^+e^-\to Y(4660)\,({\mbox  or } \,\gamma^\ast \,\mbox{or higher resonances}) \to
Z_1(4475) \pi \to D^{(*)}\bar{D}^{(*)} \pi$ is a suitable processes
to search for $Z_1(4475)$.
\section{Summary}\label{sec5}

Motivated by LHCb's confirmation of the
spin parity of $Z_1(4475)$ \cite{Aaij:2014jqa},
we have investigated the implications of this measurement. Under the
molecular assumption, we have considered three cases: (1)
$Z_1(4475)$ as the P-wave excitation of the S-wave $D_1 {\bar
D}^\ast$ or $D_2 {\bar D}^\ast$ molecule, (2) $Z_1(4475)$ as the
S-wave molecule composed of a $D$ or $D^*$ meson and a D-wave vector
$D$ meson, or (3) $Z_1(4475)$ as the cousin molecular state of
$Z_c(3900)$ and $Z_c(4020)$ composed of a $D$ or $D^*$ meson and
their radial excitations. With the help of the heavy quark symmetry,
we have studied the radiative and strong decay patterns of
$Z_1(4475)$.

In the heavy quark symmetry limit, the S-wave molecule composed of a
$D$ or $D^*$ meson and a D-wave vector $D$ meson does not decay into
the $\psi'\pi$ final state, which is the discovery mode of
$Z_1(4475)$. Therefore, this molecule scheme seems unfavorable.

If $Z_1(4475)$ could be the P-wave excitation of the $D_1 {\bar
D}^\ast$ or $D_2 {\bar D}^\ast$ molecule, we can study their
radiative and strong decay patterns together with their S-wave
molecular ground states simultaneously, the radiative decays of
which are presented in Ref. \cite{Ma:2014ofa}. Within this scheme, the
non-observation of $Z_1(4475)$ in the $J/\psi\pi$ mode is always a
serious challenge. There exists no manifest symmetry forbidding this
mode. The same challenge holds for the tetraquark interpretation.
Moreover, if $Z_1(4475)$ is the P-wave molecule, where is the ground
state? Is it $Z_0(4239)$? If so, can we find a natural framework to
explain its large binding energy around 190 MeV?

If $Z_1(4475)$ happens to be the  molecular cousin of $Z_c(3900)$
and $Z_c(4020)$ composed of a $D$ or $D^*$ meson and their radial
excitations, it decays into $J/\psi\pi$ and $\psi'\pi$ easily.
However, it will not decay into $\psi(1^3D_1)\pi$ in the heavy quark
symmetry limit. The neutral component will also decay into
$\chi_{cJ}$ through the $M1$ transition. The resulting decay width
ratio is 1:3:5. Since $Z_1(4475)$ contains one radial excitation as
its molecular component, one may expect that $Z_1(4475)$ may decay
into the final state containing a radial excitation more easily. Up
to now, the puzzling charged state $Z_1(4475)$ remains very
mysterious. Hopefully the present work will help us to understand
its underlying structure better.

\section*{Acknowledgments}

This project is supported by the National Natural Science Foundation
of China under Grants No. 11222547, No. 11175073, No. 11035006, No.
11375240 and No. 11261130311, the Ministry of Education of China
(FANEDD under Grant No. 200924, SRFDP under Grant No. 2012021111000,
and NCET), the China Postdoctoral Science Foundation under Grant No.
2013M530461, and the Fok Ying Tung Education Foundation (Grant No.
131006).

\end{document}